\documentclass[color]{aib}

\usepackage[latin1]{inputenc}

\AIBtitle{{\rwthblue The \name{} Protocol: Securely Outsourcing Sensor Data to the Cloud}}
\AIBauthors{Martin Henze, Ren{\'e} Hummen, Roman Matzutt, Klaus Wehrle}
\AIBnumber{2016}{06}
\AIBdate{July 2016}

\usepackage{cite}
\usepackage{paralist}
\usepackage{xcolor}

\newcommand{\name}{SensorCloud}
\renewcommand{\figurename}{Figure}

\begin{document}

\makeAIBtitle

\title{The \name{} Protocol: Securely\\ Outsourcing Sensor Data to the Cloud}

\author{Martin Henze, Ren{\'e} Hummen, Roman Matzutt, Klaus Wehrle}

\institute{
  Communication and Distributed Systems\\
  RWTH Aachen University, Germany\\
  Email: \email{$\{$henze, hummen, matzutt, wehrle$\}$@comsys.rwth-aachen.de}  
}
 
\date{\today}

\maketitle

\pagestyle{empty}

\begin{abstract}
The increasing deployment of sensor networks, ranging from home networks to industrial automation, leads to a similarly growing demand for storing and processing the collected sensor data.
To satisfy this demand, the most promising approach to date is the utilization of the dynamically scalable, on-demand resources made available via the cloud computing paradigm.
However, prevalent security and privacy concerns are a huge obstacle for the outsourcing of sensor data to the cloud.
Hence, sensor data needs to be secured properly before it can be outsourced to the cloud.

When securing the outsourcing of sensor data to the cloud, one important challenge lies in the representation of sensor data and the choice of security measures applied to it.
In this paper, we present the \name{} protocol, which enables the representation of sensor data and actuator commands using JSON as well as the encoding of the object security mechanisms applied to a given sensor data item.
Notably, we solely utilize mechanisms that have been or currently are in the process of being standardized at the IETF to aid the wide applicability of our approach.
\end{abstract}

%
%

\section{Introduction}

Recent advances in ubiquitous computing and wireless sensor networks continue to obliterate the boundaries between the physical and the digital world~\cite{akyildiz_sensor-survey_2002,hummen_sensorcloud_2012,henze_cps_2016,henze_ipacs_2014}.
Sensor networks can be utilized in a large variety of deployments, ranging from personal homes over offices and cars to industrial facilities and public areas~\cite{eggert_sensorcloud_2014,henze_sensorcloud_2014}.
To cope with the resulting increase in demands for storing and processing sensor data, cloud computing elastically provides the necessary computation and storage resources~\cite{henze_sensorcloud_2014}.
Cloud computing allows the collection, processing, and storage of sensor data at large scales and as well enables the world-wide sharing of said data~\cite{henze_scslib_2014,henze_sensorcloud_2013}. 
In the context of the \name{} project~\cite{eggert_sensorcloud_2014}, we consider a scenario in which operators of sensor networks (i.e., private users, companies, or public institutions) connect their sensor networks to the cloud~\cite{henze_sensorcloud_2014}, where collected sensor data is processed by cloud services selected by the sensor network operator~\cite{hummen_sensorcloud_2012,eggert_sensorcloud_2014,henze_sensorcloud_2013,henze_sensorcloud_2014}.
Besides the remarkable advantages of cloud computing, it is important to note that sensor data often contains sensitive information.
Hence, when transferring this sensitive data to entities outside of trusted sensor networks, it might, e.g., be unintentionally forwarded to third parties or used for non-authorized purposes~\cite{henze_cloudannotations_2013,henze_cloud-data-handling_2013,henze_ipacs_2016,pearson_privacy_2010}.
Furthermore, data stored and processed in the cloud might be subject to access by the cloud provider or governmental agencies~\cite{henze_privercloud_2016}.
Thus, one major challenge when interconnecting sensor networks with the cloud is to account for the aforementioned security and privacy concerns.

As part of the efforts of the \name{} project, we developed a trust point-based security architecture for outsourcing sensor data to the cloud and the \name{} security library~\cite{eggert_sensorcloud_2014,hummen_sensorcloud_2012,henze_sensorcloud_2013,henze_sensorcloud_2014,henze_scslib_2014}.
One important challenge in securely outsourcing sensor data to the cloud lies in the representation of sensor data and the corresponding security measures taken to protect the sensor data.
In this paper, we report on the \name{} protocol, which has jointly been developed within the \name{} project to represent sensor data and actuator commands using JSON and subsequently secure this data using object security mechanisms.
To this end, we rely on mechanisms that have been or currently are in the process of being standardized at the IETF and provide a best practice on how to utilize and combine them in an actual system.

The remainder of this document is structured as follows.
In Section~\ref{sec:scenario}, we present the \name{} scenario in more detail and provide references to more detailed descriptions of the overall security architecture.
Section~\ref{sec:json-representation} defines the JSON-based representation of sensor data and actuator commands in \name{}.
In Section~\ref{sec:security-extensions}, we describe the security extensions to this representation to realize the secure outsourcing of sensor data to the cloud.
We conclude this paper in Section~\ref{sec:conclusion}.

\section{\name{} Scenario}\label{sec:scenario}

\begin{figure}[t]
\centering
\includegraphics{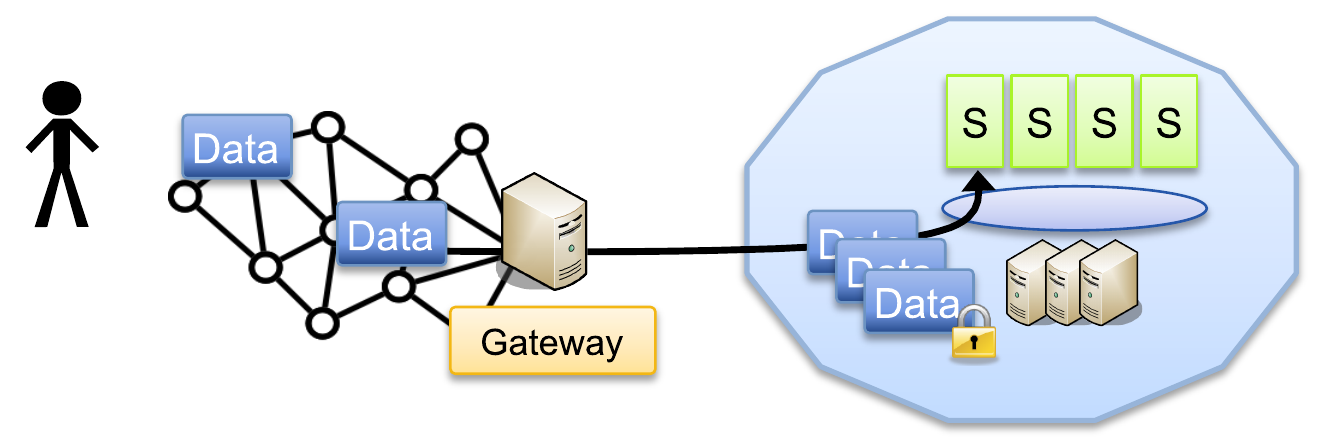}
\caption{In the \name{} scenario, data flows from sensor networks through a dedicated gateway to the cloud. There, the data is stored securely and can only be accessed by authorized services.}
\label{fig:scenario}
\end{figure}

In \name{}, we consider a scenario where each sensor network (with an arbitrary number of sensor nodes) is connected to the cloud via a dedicated gateway as depicted in Figure~\ref{fig:scenario}.
Our goal is to store data securely in the cloud such that it can only be processed by authorized cloud services.
To this end, the gateway encrypts sensitive sensor readings using a symmetric cipher before uploading it to the cloud.
The encryption process is influenced by a user-configurable access control list containing services that are authorized to (partially) obtain and process the user's sensor data.
Now, only entities in possession of the symmetric key used for encrypting a piece of sensor data (referred to as a data item) have access to this specific data item.
Hence, to grant a cloud service access to a given data item the gateway has to provide this cloud service with the corresponding key.
To this end, the gateway asymmetrically encrypts the corresponding symmetric key with the public key of the cloud service that should gain access to the sensor data and forwards the resulting encrypted key to the respective cloud service.
More details, especially with respect to the design and implementation of the underlying security architecture as well as our choice of cryptographic primitives, can be found in the corresponding publications~\cite{hummen_sensorcloud_2012,henze_sensorcloud_2013,henze_sensorcloud_2014}.

Notably, sensor data originating from a single (possibly virtual) sensor node can contain multiple sensor readings from different sensors.
For example, one data item measured by a meteorological sensor might consist of multiple single sensor readings such as humidity and temperature.
Hence, \name{} supports the transmission of multiple sensor readings of one sensor node in a single message~\cite{henze_sensorcloud_2014,henze_scslib_2014}.
As a cloud service might only be granted access to parts of the sensor readings, \name{} supports the encryption of individual parts of sensor data, thus realizing fine-grained access control.

The processing of sensor data by a cloud service requires the verification of the integrity of received data, the decryption of the symmetric key, and finally the decryption of the actual sensor data.
These operations have been implemented in the \name{} security library, which allows the transparent decryption of sensor data and the verification of its integrity by a cloud service~\cite{henze_scslib_2014}.
It is available as open source software under the MIT license\footnote{\texttt{https://code.comsys.rwth-aachen.de/redmine/projects/scslib}}.

\section{\name{} JSON Representation}\label{sec:json-representation}

The remainder of this document defines the JSON-based messages used by the \name{} protocol to encode data items as well as additional configuration messages.
More precisely, the JSON-based message layout defined by this document is used for both internal communication within the gateway and cloud, respectively, and for communication between those entities.

\subsection{Message Definition}\label{sec:message-definition}

The JSON message header MUST be included for transmissions between the gateway and the cloud. 
It MAY be omitted for internal data exchanges between components.

\begin{verbatim}
{
    "ver":"<number>", /* specification version */
    "seq":"<number>", /* for message acknowledgments */
    "pl":"[<messages>]" /* actual message payload */
}
\end{verbatim}

The field \texttt{ver} contains the version number of the protocol, which defines the structure of the remainder of the message. 
Hence, whenever the message structure or encapsulated payloads are subject to changes, the version number MUST be increased.
The receiving peer MUST support the indicated version number.
Otherwise, it SHOULD notify the sending peer that the message was dropped. 
The version number is a positive integer and is currently defined as \texttt{1}.

The sequence number \texttt{seq} is currently unused and MUST therefore be set to \texttt{0}.
However, it enables end-to-end acknowledgments at the application layer.
The corresponding retransmission mechanism will need to be developed in the future.
If used, \texttt{seq} is a positive integer that MUST be increased by one for each JSON message header sent. 
A potential wrap-around of the sequence number is to be expected and implementations MUST interpret this case as an incrementation of the sequence number.

The messages to be transmitted using one header are stored within an array in the field \texttt{pl}.
Section~\ref{sec:payload-types} defines a list of message types supported by \name{}.
Additional message types can be defined as necessary.
Messages MUST contain the field \texttt{typ} to indicate their message type for message processing purposes.
Moreover, they MUST include the \texttt{gw} field to indicate the gateway device that participates in the communication.
Among other aspects, this information is required for applying object security measures.
Note that multiple messages of different types can be batched using one message header in order to reduce the overall communication overhead.

\subsection{Defined Message Types}\label{sec:payload-types}
Each message indicates its message type via the field \texttt{typ}, which MUST be included in any message.
The message type is a positive integer value and can currently take one of the values described in Table~\ref{tab:msg-types}.

\setcounter{table}{-1}
\begin{table}
\begin{longtable}[c]{|p{2cm}p{6cm}|}
\hline
\centering{Value} & \centering{Semantics}\tabularnewline
\hline
\centering{1} & \centering{Sensor Data Message}\tabularnewline
\centering{2} & \centering{Sensor Data Request}\tabularnewline
\centering{3} & \centering{Configuration Message}\tabularnewline
\centering{4} & \centering{Actuator Command}\tabularnewline
\centering{5} & \centering{Actuator Response}\tabularnewline
\centering{400} & \centering{Data Key Upload}\tabularnewline
\centering{401} & \centering{Data Key Download}\tabularnewline
\centering{402} & \centering{Public Key Download}\tabularnewline
\centering{403} & \centering{Public Key Response}\tabularnewline
\hline
\end{longtable}
\vspace{0.5em}
\caption{Message types defined for the \name{} protocol.}
\label{tab:msg-types}
\end{table}

The remainder of this section defines the structure of messages of the respective types in detail.
Section~\ref{sec:sensor-data-definition} describes messages carrying sensor data while Section~\ref{sec:requesting-sensor-data} defines sensor data queries to be sent by cloud services.
Section~\ref{sec:configuration-definition} defines configuration messages sent to describe the layout of sensor data originating from single sensor nodes.
Afterwards, Section~\ref{sec:actuator-command-definition} and Section~\ref{sec:actuator-response-definition} describe commands to be sent to sensor nodes with actuation capabilities and the format of responses to such commands, respectively.
We finally describe messages related to the management of cryptographic keys as a part of the security extensions in Section~\ref{sec:key-management}.

\subsection{Transmitting Sensor Data}\label{sec:sensor-data-definition}

For sensor data, the JSON representation is strongly based on the JSON definitions of the Media Types for Sensor Markup Language (SenML) \cite{jennings_senml_2016}. 
SenML is currently being standardized at the IETF. 
Only a short example is discussed here to show how SenML integrates into the sensor data representation of \name{}. 
For specifics about the SenML representation, reading Section 6 of the corresponding SenML draft \cite{jennings_senml_2016} is strongly encouraged.

\begin{verbatim}
{
    "typ":"1", /* type of the message payload */
    "gw":"<string>", /* unique gateway ID */
    "bn":"<string>", /* sensor device ID */
    "bt":"<number>", /* base time of sensor readings */
    "e":[{
        "n":"<string>", /* sensor ID */
        "t":"<number>", /* optional time value */
        "sv":"<string>" /* sensed value as string */
    }]
}
\end{verbatim}

Each message encodes one data item, i.e., one piece of sensor data as described in Section~\ref{sec:scenario}.
Each data item starts with the field \texttt{typ}, which is always set to \texttt{1}, indicating that the message contains sensor data to be uploaded.
The fields \texttt{src} and \texttt{bn} contain the unique identifier of the processing gateway and the identifier of the originating sensor device, respectively.
The sensor device identifier \texttt{bn}  thereby MUST be unique on a per-gateway basis. 
Data items also carry a timestamp \texttt{bt}, a positive integer denoting the time (in milliseconds since UNIX epoch) at which the oldest sensor reading within the data item was measured.

Sensor data from a single (virtual) sensor device can consist of readings from multiple sensors of the same device (e.g., humidity and temperature), a series of sensor readings from the same sensor, or a combination of the two.
Each sensor reading is encoded as a JSON object in the array \texttt{e}.
The name field \texttt{n} thereby allows to identify the specific sensor, whereas \texttt{t} allows to identify the time offset (in milliseconds) of a sensor reading relative to the base time \texttt{bt}.
Objects in the array \texttt{e} MUST be ordered according to an ascending alphanumeric order with respect to the field \texttt{n}.
If \texttt{n} is equal for multiple sensor readings, these elements MUST be ordered in ascending order of the time value \texttt{t}.
The array \texttt{e} MUST NOT contain two or more elements with the same combination of \texttt{n} and \texttt{t} values.
Hence, the granularity of \texttt{t} MUST be sufficient to guarantee this property.
In \name{}, the granularity of \texttt{t} is milliseconds. 
Moreover, the individual JSON identifiers MUST appear in the order shown above.
This order is required to preserve the payload structure to enable cryptographically verifying the data item's integrity later on.

The actual sensed data is stored in the field \texttt{sv} (string value).
SenML already defines a number of primitive data types and \texttt{sv} is only one of them.
Other data types such as \texttt{bv} (boolean value) may be used as well, but are currently not supported by \name{}.
However, using different data types can aid casting of the sensed value in typed programming languages.

\subsection{Requesting Sensor Data}\label{sec:requesting-sensor-data}

Third-party services can query the cloud for sets of data items to process them.
The cloud, holding a replica of the data owner's access control list (cf. Section~\ref{sec:scenario}), then returns the data items that are specified in the query and accessible by the service (to unburden the service from attempting to decrypt unaccessible data items).
Data item queries have the following structure, which is based on the structure of data items:

\begin{verbatim}
{
    "typ":"2", /* payload is a sensor data request */
    "gw":"<string>", /* unique gateway ID */
    "srv":"<string>", /* unique service ID */
    "lim":"<number>", /* response length limit */
    "off":"<number>", /* offset in specified sensor data stream */
    "bt":["<number>"], /* base time of sensor readings */
    "bn":["<string>"], /* sensor device IDs */
    "e":[{
        "n":"<string>", /* sensor IDs */
    }]
}
\end{verbatim}

Sensor data requests always have a message type \texttt{typ} of \texttt{2}.
The fields \texttt{srv} and \texttt{gw} specify the service requesting sensor data and the gateway responsible for the queried sensor data, respectively.
The fields \texttt{lim} (limit) and \texttt{off} (offset) MAY be used by the requesting service to further control the query.
If \texttt{lim} is specified, the cloud MUST NOT return more than the specified number of data items in the response.
If \texttt{lim} is not given, the cloud SHOULD return the whole response, i.e., all currently available data items matching the query.
Furthermore, the cloud MUST ignore the first \texttt{off}-many data items it would include in the response.
If \texttt{off} is not given, its default value is \texttt{0}.

The remaining fields define the query itself.
The base time array \texttt{bt} specifies a time range for the creation times of data items (as given by the field \texttt{bt}, c.f. Section~\ref{sec:sensor-data-definition}) to be included in the response.
The maximum length of \texttt{bt} is two; in this case, the first element specifies the lower time bound and the second element specifies the upper time bound in milliseconds elapsed since UNIX epoch, respectively.
Note that both bounds are including, i.e., data items to be included in the response have a value of \texttt{bt} that is larger than or equal to the lower bound and smaller than or equal to the upper bound, respectively.
If only one element is given, it is treated as the lower time bound for the data item creation time.
If \texttt{bt} is empty or not given, all data items matching by other criteria are to be returned.
The array \texttt{bn} specifies a set of identifiers of sensor nodes whose sensor readings shall be processed by the querying cloud service.
For instance, the affected sensor nodes can be derived from the cloud's copy of the data owner's access control list.
However, the exact specification of how services obtain the required sensor node identifiers is out of the scope of the \name{} protocol.
The array \texttt{e} contains JSON objects each containing a sensor identifier.
The cloud MUST return only those data items that have been sensed by the sensors specified in the array \texttt{e}.
In summary, when receiving a query the cloud MUST return exactly those data items that
\begin{inparaenum}[(a)]
\item{have been created within the timespan given by \texttt{bt},}
\item{originate from a sensor node given in \texttt{bn}, and}
\item{contain measurements by sensors of a type specified in the array \texttt{e}.}
\end{inparaenum}

\subsection{Configuration Definition}\label{sec:configuration-definition}

\name{} uses dedicated configuration messages to define the layout of sensor data readings emitted by a specific sensor device.
This way, sensor networks consisting of devices with heterogeneous capabilities can be accounted for.
The structure of configuration messages is as follows:

\begin{verbatim}
{
    "typ":"3", /* payload is a configuration message */
    "gw":"<string>", /* unique gateway ID */
    "bn":"<string>", /* sensor device ID */
    "js":"<string>", /* JSON schema */
}
\end{verbatim}

Configuration messages always use the message type \texttt{3} in the field \texttt{typ}.
The fields \texttt{gw} and \texttt{bn} contain the unique identifier of the processing gateway and the identifier of the addressed sensor device, respectively. 
The field \texttt{js} contains the JSON schema specifying the structure of sensor data read by the sensor device that is uniquely identified by the combination of \texttt{gw} and \texttt{bn}.

\subsection{Actuator Command Definition}\label{sec:actuator-command-definition}

Sensor network devices can also be equipped with actuators, allowing authorized entities, e.g., cloud services authorized by the sensor network owner, to externally trigger certain actions based on current sensor measurements.
In \name{}, actuator messages are used to forward actuator commands.
Their structure is based on the structure of sensor data payloads and looks as follows:

\begin{verbatim}
{
    "typ":"4", /* type of the message payload */
    "gw":"<string>", /* unique gateway ID */
    "srv":"<string>", /* unique service ID */
    "bn":"<string>", /* actuator device ID */
    "seq":"<number>", /* Sequence number (optional) */
    "fn":"<string>", /* Function name (optional) */
    "e":[{
        "n":"<string>", /* Parameter key */
        "sv":"<string>" /* Parameter value */
    }]
}
\end{verbatim}

Actuator command messages use a message type \texttt{typ} of \texttt{4}.
The field \texttt{gw} specifies the destination gateway for the actuator command.
The destination gateway decides whether or not to forward an actuator command based on internal access control lists containing those cloud services that have been authorized by the data owner to access fractions of her sensor data.
The field \texttt{srv} contains a unique identifier of the triggering cloud service to allow the gateway to match the actuator command to an entry in the access control list.
The actuator device to execute the command can be uniquely identified by the destination gateway by considering the field \texttt{bn}.
Optionally, an actuator command can also carry a sequence number \texttt{seq}, which is used by cloud services if a response to the command is expected, e.g., an acknowledgment.
In this case, the cloud service MUST ensure that \texttt{seq} is chosen in a way that the mapping between actuator commands sent and responses received is unambiguous.
The optional field \texttt{fn} MAY be used to specify the name or identifier of a function to be called at the gateway for RPC-like interactions \cite{birrell_rpc_1984}.
The array \texttt{e} contains a set of parameters in the form of key-value pairs (\texttt{n} as key, \texttt{sv} as value) that shall be passed to the called function. 
If \texttt{fn} is not included, the default behavior is to set the parameters given by the fields \texttt{n} in \texttt{e} at the actuator \texttt{bn} to the respective values \texttt{sv}.

Each actuator command message may only contain a single actuator command.
If multiple actuator commands are to be initiated, multiple messages of type \texttt{4} can be batched using a single message header, which was defined in Section~\ref{sec:message-definition}.

\subsection{Actuator Response Definition}\label{sec:actuator-response-definition}

Actuator commands in \name{} can trigger actuator responses, e.g., acknowledging a successful execution of the triggered action.
In such a case, an actuator response is sent to the originating cloud service, thereby enabling RPC-like interactions \cite{birrell_rpc_1984}. 
Note that the sequence number \texttt{seq} MUST be set in the corresponding actuator command in order to enable the receiving service to match the response to the corresponding request.
The structure of an actuator response message is as follows, in analogy to the structure of an actuator command message:

\begin{verbatim}
{
    "typ":"5", /* type of the message payload */
    "gw":"<string>", /* unique gateway ID */
    "srv":"<string>", /* unique service ID */
    "bn":"<string>", /* actuator device ID */
    "seq":"<string>", /* Sequence number */
    "fn":"<string>", /* Function name (optional) */
    "e":[{
        "n":"<string>", /* name */
        "sv":"<string>" /* value as string */
    }]
}
\end{verbatim}

To denote an actuator response message, its message type \texttt{typ} has the fixed value of \texttt{5}.
The fields \texttt{gw} and \texttt{srv} denote the unique identifiers of the gateway and cloud service involved in the actuator command, respectively.
These fields are copied from the corresponding actuator command message.
Additionally, the fields \texttt{bn}, \texttt{seq}, and \texttt{fn} are copied from the actuator command message to enable the cloud service receiving the response to map it to any pending state it holds for the actuator command.
If \texttt{fn} is not included in the original actuator command message, it is also not included in the response and the service assumes that the actuator responds to only setting the parameter values specified in the actuator command message.
The array \texttt{e} contains a (possibly empty)  set of return values from the previously called function. 
For instance, this array can contain an error message (e.g., \texttt{n} is set to \texttt{"err"} and \texttt{sv} contains an error code or error message).

Each actuator command response may only refer to a single actuator command.
Analogously to actuator command messages, multiple actuator command responses can be batched using a single message header as defined in Section~\ref{sec:message-definition}.

\section{Security Extensions}\label{sec:security-extensions}

\name{} bases the protection of sensitive sensor readings on specifications of the JOSE WG at the IETF. 
More precisely, JSON Web Encryption (JWE) \cite{jones_jwe_2015} is employed to encrypt sensitive sensor information, whereas integrity protection and authentication for the complete data item is provided via JSON Web Signature (JWS) \cite{jones_jws_2015}.
For specifics, reading the respective specifications~\cite{jones_jwe_2015,jones_jws_2015} is strongly encouraged.
In the following, we first describe the integration of JWE into the \name{} protocol in Section~\ref{sec:encryption}, then Section~\ref{sec:signing} documents how JWS is used in this protocol, and finally Section~\ref{sec:key-management} describes the key management in \name{}.

\subsection{Encryption}\label{sec:encryption}

If values are to be encrypted, JWE extends the data item with information about the ciphersuite used as specified by the JWE JSON Serialization.
Furthermore, any binary data occurring (e.g., ciphertexts or initialization vectors) are encoded using the \texttt{base64url} encoding.
Four values are represented in a JWE: the header, initialization vector, ciphertext, and authentication tag. 
In \name{}, JWE-encrypted JSON structures are encapsulated in a JSON array  \texttt{ev} (encrypted value), replacing the field \texttt{sv} of an unencrypted value:

\begin{verbatim}
"ev":[{
    "unprotected":{
        "alg":"dir",
        "enc":"AESGCM256",
        "kid":"<string>",
        "typ":"<string>"
    },
    "iv":"<initialization vector (base64url-encoded)>",
    "ciphertext":"<ciphertext (base64url-encoded)>",
    "tag":"<authentication tag (base64url-encoded)>"
}]
\end{verbatim}

The JWE header, stored in the field \texttt{unprotected}, contains the fields \texttt{alg}, \texttt{enc}, \texttt{kid}, and \texttt{typ}. 
This field is not covered by the authentication tag and therefore it does not need to be \texttt{base64url}-encoded.
For \name{}, we fix the value of the \texttt{alg} (algorithm) field to \texttt{dir} (direct), indicating that the symmetric key used to decrypt the ciphertext is not encrypted asymmetrically itself, i.e., the given keying material is passed directly to the ciphersuite given by the field \texttt{enc}.
The ciphersuite is fixed as \texttt{AESGCM256}, i.e., AES with keys of length 256 bit using the Galois Counter Mode (GCM).
The field \texttt{kid} (key identifier) is used to store the identifier of the key used for encryption, i.e., its SHA-1 hash value.
The field \texttt{typ} encodes the data type of the encrypted value, e.g., \texttt{sv} in \name{}.
The initialization vector used during the symmetric encryption is stored in the field \texttt{iv}, whereas \texttt{tag} holds the authentication tag, which is optionally used depending on the AES mode of operation, e.g., GCM.
Finally, \texttt{ciphertext} contains the encrypted value itself.
For example, if one sensor reading is to be encrypted and one is not, this gives the following representation:

\begin{verbatim}
{
    "typ":"1",
    "gw":"<string>",
    "bn":"<string>",
    "bt":"<number>",
    "e":[
    {
        "n":"<string>",
        "ev":[
        {
            "unprotected":
            {
                "alg":"dir",
                "enc":"AESGCM256",
                "kid":"<string>",
                "typ":"sv"
            },
            "iv":"<string (base64url-encoded)>",
            "ciphertext":"<string (base64url-encoded)>",
            "tag":"<string (base64url-encoded)>"
        }]
    },
    {
        "n":"<string>",
        "sv":"<string>"
    }]
}
\end{verbatim}

After decryption, the field identifier \texttt{ev} is replaced with the \texttt{typ} given in the \texttt{unprotected} header, i.e., \texttt{sv} in \name{} and the field's content is replaced with the decrypted value.
Then, the data item can be processed normally. 

\subsubsection{Advanced Encryption Schemes.}\label{sec:advanced-use-cases-for-encryption}

\name{} allows for two more advanced schemes for the encryption of sensor data: encrypting the whole measurement array of one data item and packing the encryptions of multiple values into one JWE object.

The definition of \texttt{ev} also allows to encrypt the array \texttt{e} in the sensor data payload in its entirety instead of on a per-sensor-reading basis.
In this case, the plaintext is composed of a canonical serialization of the array \texttt{e} and the JWE object \texttt{ev} replaces the array \texttt{e}.
This reduces the encryption overhead in comparison to the case where each sensor value is encrypted separately. 
However, the trade-off is that information about the individual sensor readings, e.g., the sensor identifier \texttt{n}, is encrypted as well, hence selectively processing only subsets of the sensed data is not possible anymore without decrypting the whole array.
Thus, the encryption of the entire \texttt{e} array is discouraged.

Additionally, the use of \texttt{ev} is not restricted to only encrypting single sensor readings. 
Instead, the definition also allows for multiple fields to be encrypted and encapsulated in a single \texttt{ev} array if they are at the same JSON hierarchy level.
In \name{}, the encryption of non-sensor-value fields (e.g. \texttt{gw}, \texttt{bt}) within a data item is discouraged because they contain important information necessary for indexing of sensor data in the cloud. 
However, configuration messages  may contain fields more suited for encryption.
To further clarify the use of \texttt{ev} with multiple sensitive fields at the same hierarchy level, assume that a configuration message is extended with an additional field \texttt{ref}:

\begin{verbatim}
{  
    "typ":"3",
    "gw":"<string>",
    "bn":"<string>",
    "ref":"<string>", /* reference ID */
    "js":"<string>",  /* JSON schema */
}
\end{verbatim}

Further assume that both fields \texttt{ref} and \texttt{js} contain sensitive data and must therefore be encrypted. 
This yields the following encrypted representation, where both encrypted values are packed into the same array \texttt{ev}:

\begin{verbatim}
{
    "typ":"3",
    "gw":"<string>",
    "bn":"<string>",
    "ev":[
    {
        "unprotected":
        {
            "alg":"dir",
            "enc":"AESGCM256",
            "kid":"<string>",
            "typ":"ref"
        },
        "iv":"<string (base64url-encoded)>",
        "ciphertext":"<string (base64url-encoded)>",
        "tag":"<string (base64url-encoded)>"
    },
    {
        "unprotected":
        {
            "alg":"dir",
            "enc":"AESGCM256",
            "kid":"<string>",
            "typ":"js"
        },
        "iv":"<string (base64url-encoded)>",
        "ciphertext":"<string (base64url-encoded)>",
        "tag":"<string (base64url-encoded)>"
    }]
}
\end{verbatim}

In addition to the case mentioned above, a message can also contain multiple \texttt{ev} arrays as long as they do not fall into the same scope, e.g., they are not on the same hierarchy level or a certain value in multiple objects in an array shall be encrypted.
Each \texttt{ev} array thereby contains all encrypted fields within the same scope. 
To illustrate this, assume that sensor data messages are extended with two new fields \texttt{bp} and \texttt{bq}, both of which shall be encrypted:

\begin{verbatim}
{
    "typ":"1",
    "gw":"<string>",
    "bn":"<string>",
    "bt":"<number>",
    "bp":"<string>",
    "bq":"<string>",
    "e":[
    {
        "n":"<string>",
        "sv":"<string>"    
    },
    {
        "n":"<string>",
        "sv":"<string>"    
    }]
}
\end{verbatim}

In this case, the corresponding encrypted message looks as follows (each value \texttt{sv} is encrypted as well):

\begin{verbatim}
{
    "typ":"1",
    "gw":"<string>",
    "bn":"<string>",
    "bt":"<number>",
    "ev":[{... /* encryption of bp */},{... /* encr. of bq */}],
    "e":[
    {
        "n":"<string>"
        "ev":[{... /* encr. of sv */}]
    },
    {
        "n":"<string>"
        "ev":[{... /* encr. of sv */}]
    }]
}
\end{verbatim}

Note that this representation contains three \texttt{ev} fields: one at the first hierarchy level that contains the encrypted values of \texttt{bp} and \texttt{bq} as well as another two that encrypt the individual sensor readings, respectively.

\subsection{Signing}\label{sec:signing}

In order to protect message integrity, messages can be signed digitally by their respective senders.
How to digitally sign JSON objects such as \name{} messages is specified by JSON Web Signatures (JWS)~\cite{jones_jws_2015}, using the \texttt{base64url} encoding for any binary values that shall be integrity-protected and the signature itself.
By signing a message, the message is extended with a field \texttt{sig} containing a JWS header and the signature:

\begin{verbatim}
"sig":{
    "signatures":[{
        "header":{"alg":"ES256"},
        "signature":"<signature contents (base64url-encoded)>"
    }]
}
\end{verbatim}

To provide a sufficient security level until at least the year 2030~\cite{nist_keylengths_2016}, \name{} signs messages using ECDSA over the elliptic curve P-256 and SHA-256 as the corresponding hash function.
This is indicated by the \texttt{header} parameter \texttt{alg}, which is set to \texttt{"ES256"}.
Due to the \name{} limitation to a single signature per message, we limit the length of the signatures array to one element.
The key used for signing can be derived from the gateway identifier \texttt{gw} given in the message to be signed. 
If no \texttt{gw} field exists, the \texttt{header} MUST additionally include a field \texttt{kid} containing the gateway identifier.

By signing each message individually, the receiver of a message is able to verify each message individually after transmission.
Furthermore, messages are always signed as a whole.
To ensure consistency for signing a message, an empty object \texttt{sig} is appended to the message prior to signing.
Additionally, the message is canonicalized following the recommendations of Canonical JSON~\cite{olpc_canonical}. 
Hence, a message may look as follows before the signature algorithm is applied:

\begin{verbatim}
{
    "typ":"1",
    "gw":"<string>",
    "bn":"<string>",
    "bt":"<number>",
    "e":[
    {
        "n":"<string>",
        "ev":[
        {
            "unprotected":
            {
                "alg":"dir",
                "enc":"AESGCM256",
                "kid":"<string>",
                "typ":"<string>"
            },
            "iv":"<string (base64url-encoded)>",
            "ciphertext":"<string (base64url-encoded)>",
            "tag":"<string (base64url-encoded)>"
        }]
    }],
    "sig":{}
}
\end{verbatim}

Then, the signature is computed over the \texttt{base64url}-encoded SHA-256 hash value of the message.
Finally, the field \texttt{sig} is filled by including the JWS header and the signature value as the single element of the \texttt{signatures} array:

\begin{verbatim}
{
    "typ":"1",
    "gw":"<string>",
    "bn":"<string>",
    "bt":"<number>",
    "e":[
    {
        "n":"<string>",
        "ev":[
        {
            "unprotected":
            {
                "alg":"dir",
                "enc":"AESGCM256",
                "kid":"<string>",
                "typ":"<string>"
            },
            "iv":"<string (base64url-encoded)>",
            "ciphertext":"<string (base64url-encoded)>",
            "tag":"<string (base64url-encoded)>"
        }]
    }],
    "sig":{
        "signatures":[{
            "header":{"alg":"ES256"},
            "signature":"<string (base64url-encoded)>"
        }]
    }
}
\end{verbatim}

As a result of this approach, multiple sensor readings SHOULD be split into several messages if they are not to be used or stored together.
Otherwise, all sensor readings that previously have been retrieved from the same message have to be collected again to verify a message.

\subsection{Key Management}\label{sec:key-management}

In this section, we describe the key management within \name{}.
First, we define the messages sent for uploading and downloading data keys, respectively.
In \name{}, a data key refers to a symmetric cryptographic key used to encrypt a series of data items.
Then, we argue why the \name{} protocol does not define a dedicated message to upload public keys of participating entities, i.e., gateways and cloud services.
Finally, we define the message to be sent when a participating entity has to download the public key of another entity.

\subsubsection{Data Key Upload.}\label{sec:datakey-upload}

Data keys are symmetric keys used to encrypt and decrypt data items.
In \name{}, gateways periodically exchange the data keys used for different streams of sensor data.
In order to enable cloud services to process the uploaded sensor data, gateways need to share data keys with those cloud services that are authorized to access this sensor data.
Therefore, for each service authorized to read some sensor data a gateway uploads copies of the needed data keys, encrypted with the respective service's public key, to the cloud.
The structure of messages for the upload of a data key is as follows:

\begin{verbatim}
{
    "typ":"400", /* message type */
    "gw":"<string>", /* gateway ID */
    "srv":"<string>", /* service ID */
    "bt":["<number>", "<number>"], /* key validity timerange */
    "bn":"<string>", /* sensor node ID */
    "e":[{
        "n":"<string>", /* sensor ID */
        "kid":"<string>", /* key ID */
        "k":"<string (base64url-encoded)>" /* key material */
    ]}
}
\end{verbatim}

Upload messages of data keys always have a message type \texttt{typ} of \texttt{400}.
The fields \texttt{gw} and \texttt{srv} hold the unique identifiers of the gateway uploading the encrypted data key and the cloud service that is able to obtain the key, respectively.
The array \texttt{bt} specifies a validity timespan for the data key and MUST be of length two, where the first element is the lower bound of the key's validity and the second element is the upper bound, respectively.
Both times are given as milliseconds since UNIX epoch and the timespan includes the upper and lower bounds.
The gateway SHOULD NOT encrypt data items using a data key that is already expired.
The field \texttt{bn} contains the identifier of a sensor node for which new data keys are to be uploaded.
In the array \texttt{e}, a new data key for each sensor \texttt{n} on the sensor node referred to by \texttt{bn} can be given by specifying the key's identifier, i.e., its SHA-1 hash value, in the field \texttt{kid} and the keying material (in \texttt{base64url}-encoded form) in the field \texttt{k}.
Single sensors of a sensor node MAY be omitted from the array \texttt{e}, e.g., if the readings of different sensors are subject to different security requirements and therefore use data keys of differing validity periods.

\subsubsection{Data Key Download.}\label{sec:datakey-download}

In order to be able to decrypt and then process data items, cloud services must first obtain the respective data keys.
To this extend, cloud services send messages of the following structure to the cloud:\\

\begin{verbatim}
{
    "typ":"401", /* message type */
    "gw":"<string>", /* gateway ID */
    "srv":"<string>", /* service ID */
    "kid":"<string>", /* key ID */
}
\end{verbatim}

The message type \texttt{typ} of \texttt{401} indicates a data key download message.
In this message, the service identified by \texttt{srv} requests to download the data key identified by \texttt{kid} and managed by the gateway identified by \texttt{gw} from the cloud.
The service includes its own identifier into the message in order to enable the cloud to respond with the data key specifically encrypted for the requesting service.
Furthermore, the cloud responds with a message of type \texttt{400} as specified in the previous section.

\subsubsection{Public Key Upload.}\label{sec:pubkey-upload}

\name{} does not feature a designated upload message for public keys in the \name{} protocol.
Instead, asymmetric key pairs, which determine an entity's identity, are expected to be issued by the cloud.
For instance, the cloud MAY require gateways and services to be registered by other means, e.g., via a web front-end, and issue a client certificate upon successful registration.
During this process, the respective data owner or service provider locally creates a key pair and in return obtain the client certificate, which is bound to the data owner or service provider's real identity.

\subsubsection{Public Key Download.}\label{sec:pubkey-download}

In contrast to the creation of public keys, their retrieval from the cloud must be an automated process.
The corresponding message of the \name{} protocol has the following structure:

\begin{verbatim}
{
    "typ":"402", /* message type */
    "id":"<string>" /* entity ID */
}
\end{verbatim}

The message type \texttt{typ} of \texttt{402} indicates that a public key is about to be downloaded.
The requesting entity specifies the identifier of the entity of which the public key shall be retrieved in the field \texttt{id}.
On the one hand, gateways need to request services' public keys in order to provide them with the data keys needed to process sensor data.
On the other hand, services need to request a gateway's public key in order to verify the integrity of sensor data downloaded from the cloud.

The cloud then responds with the respective public key, using a message of the following structure:

\begin{verbatim}
{
    "typ":"403", /* message type */
    "key":"<string (PEM-encoded)>" /* public key */
}
\end{verbatim}

The message type \texttt{typ} of \texttt{403} indicates a message that contains a public key that is being obtained from the cloud.
Furthermore, the field \texttt{key} holds the respective entity's public key encoded using the PEM encoding~\cite{josefsson_pem_2015}.

\section{Conclusion}\label{sec:conclusion}

When securely outsourcing sensor data to the cloud, it is important to standardize the representation of sensor data as well as the encoding of the necessary security mechanisms.
In this paper, we presented the \name{} protocol, which has been developed as a joint effort in the \name{} project as a common representation for sensor data and actuator commands as well as the necessary security mechanisms using JSON.
The \name{} protocol is one important building block of our trust point-based security architecture for sensor data in the cloud and the \name{} security library which realize the secure outsourcing of sensor data to the cloud \cite{eggert_sensorcloud_2014,hummen_sensorcloud_2012,henze_sensorcloud_2013,henze_sensorcloud_2014,henze_scslib_2014}.
In our design of the \name{} protocol, we intentionally relied on approaches that have been or currently are in the process of being standardized at the IETF.
This does not only ease the wide applicability of our approach but also provides a best practice on how to utilize and combine these standardized building blocks in an actual system.

\subsection{Additional Reading}

This paper deliberately focuses on the documentation of the \name{} protocol.
Further information on the underlying security architecture has been presented in a number of scientific publications as follows.

First conceptual and prototypical considerations of the \name{} security architecture have been presented at IEEE CloudCom 2012 \cite{hummen_sensorcloud_2012}.
This publication outlines the designed security architecture and shows its feasibility through initial measurements.
Subsequently, we published an extension of our \name{} security architecture in the International Journal of Grid and High Performance Computing \cite{henze_sensorcloud_2013}.
In this work, we present the integration of the \name{} protocol into the design of our security architecture and discuss extended measurements of our prototypical implementation.
The scientific concept of our \name{} Security Library has been presented at AASNET 2014 \cite{henze_scslib_2014}.
Finally, we contributed a chapter to the edited book \emph{Trusted Cloud Computing}, where we provide a complete overview over the developed trust point-based security architecture \cite{henze_sensorcloud_2014}.

In addition, we describe the interdisciplinary approach of the \name{} project \cite{eggert_sensorcloud_2014} and discuss potential extensions to also provide privacy of outsourced sensor data \cite{henze_ipacs_2014,henze_ipacs_2016}.

\section*{Acknowledgments}
The authors would like to thank everyone who contributed to this document (see attached list of contributors).
The \name{} project was funded by the German Federal Ministry of Economic Affairs and Energy (BMWi) under project funding reference number 01MD11049. 
The responsibility for the content of this publication lies with the authors.

%
%
\section*{List of Contributors}

The following individuals contributed to this specification of the \name{} protocol:

\begin{itemize}
\item Anupam Ashish, QSC AG
\item Benjamin Assadsolimani, RWTH Aachen University
\item Daniel Catrein, QSC AG
\item Dominik Chmiel, RWTH Aachen University
\item Martin Henze, RWTH Aachen University
\item Lars Hermerschmidt, RWTH Aachen University
\item Ren{\'e} Hummen, RWTH Aachen University
\item Roman Matzutt, RWTH Aachen University
\item Antonio Navarro P{\'e}rez, RWTH Aachen University
\item Thomas Partsch, Cologne University of Applied Sciences
\item Christian R{\"o}ller, QSC AG
\item Daniel Scholz, Cologne University of Applied Sciences
\item Andre Skusa, symmedia GmbH
\end{itemize}

%
%

%
%
\cleardoublepage
\section*{Aachener Informatik-Berichte}
\newfont{\sss}{cmr10 scaled 1000}
\newfont{\bbb}{cmbx10 scaled 1000}
\sss

{\bbb This list contains all technical reports published
  during the past three years.
  A complete list of reports dating back to 1987 is available from:
\begin{center}
  \url{http://aib.informatik.rwth-aachen.de/}
\end{center}
  To obtain copies please consult the above URL or send your request
  to:
\begin{center}
  Informatik-Bibliothek, RWTH Aachen, Ahornstr.~55, 52056 Aachen,\\
  Email: \email{biblio@informatik.rwth-aachen.de }
\end{center}}\bigskip

\begin{longtable}{lp{11cm}}

2013-01 $^\ast$ &Fachgruppe Informatik:      Annual Report 2013\\
2013-02 & Michael Reke:         Modellbasierte Entwicklung automobiler Steuerungssysteme in Klein- und mittelst\"{a}ndischen Unternehmen\\
2013-03 & Markus Towara and Uwe Naumann:         A Discrete Adjoint Model for OpenFOAM\\
2013-04 & Max Sagebaum, Nicolas R. Gauger, Uwe Naumann, Johannes Lotz, and Klaus Leppkes:         Algorithmic Differentiation of a Complex C++ Code with Underlying Libraries\\
2013-05 & Andreas Rausch and Marc Sihling:         Software \& Systems Engineering Essentials 2013\\
2013-06 & Marc Brockschmidt, Byron Cook, and Carsten Fuhs:         Better termination proving through cooperation\\
2013-07 & Andr\'{e} Stollenwerk:         Ein modellbasiertes Sicherheitskonzept f\"{u}r die extrakorporale Lungenunterst\"{u}tzung\\
2013-08 & Sebastian Junges, Ulrich Loup, Florian Corzilius and Erika \'{A}brah\'{a}m:         On Gr\"{o}bner Bases in the Context of Satisfiability-Modulo-Theories Solving over the Real Numbers\\
2013-10 & Joost-Pieter Katoen, Thomas Noll, Thomas Santen, Dirk Seifert, and Hao Wu:         Performance Analysis of Computing Servers using Stochastic Petri Nets and Markov Automata\\
2013-12 & Marc Brockschmidt, Fabian Emmes, Stephan Falke, Carsten Fuhs, and J\"{u}rgen Giesl:         Alternating Runtime and Size Complexity Analysis of Integer Programs\\
2013-13 & Michael Eggert, Roger H\"{a}u\ss{}ling, Martin Henze, Lars Hermerschmidt, Ren\'{e} Hummen, Daniel Kerpen, Antonio Navarro P\'{e}rez, Bernhard Rumpe, Dirk Thi\ss{}en, and Klaus Wehrle:         SensorCloud: Towards the Interdisciplinary Development of a Trustworthy Platform for Globally Interconnected Sensors and Actuators\\
2013-14 & J\"{o}rg Brauer:         Automatic Abstraction for Bit-Vectors using Decision Procedures\\
2013-16 & Carsten Otto:         Java Program Analysis by Symbolic Execution\\
2013-19 & Florian Schmidt, David Orlea, and Klaus Wehrle:         Support for error tolerance in the Real-Time Transport Protocol\\
2013-20 & Jacob Palczynski:         Time-Continuous Behaviour Comparison Based on Abstract Models\\
2014-01 $^\ast$ &Fachgruppe Informatik:      Annual Report 2014\\
2014-02 & Daniel Merschen:         Integration und Analyse von Artefakten in der modellbasierten Entwicklung eingebetteter Software\\
2014-03 & Uwe Naumann, Klaus Leppkes, and Johannes Lotz:         dco/c++ User Guide\\
2014-04 & Namit Chaturvedi:         Languages of Infinite Traces and Deterministic Asynchronous Automata\\
2014-05 & Thomas Str\"{o}der, J\"{u}rgen Giesl, Marc Brockschmidt, Florian Frohn, Carsten Fuhs, Jera Hensel, and Peter Schneider-Kamp:         Automated Termination Analysis for Programs with Pointer Arithmetic\\
2014-06 & Esther Horbert, Germ\'{a}n Mart\'{\i}n Garc\'{\i}a, Simone Frintrop, and Bastian Leibe:         Sequence Level Salient Object Proposals for Generic Object Detection in Video\\
2014-07 & Niloofar Safiran, Johannes Lotz, and Uwe Naumann:         Algorithmic Differentiation of Numerical Methods: Second-Order Tangent and Adjoint Solvers for Systems of Parametrized Nonlinear Equations\\
2014-08 & Christina Jansen, Florian G\"{o}be, and Thomas Noll:         Generating Inductive Predicates for Symbolic Execution of Pointer-Manipulating Programs\\
2014-09 & Thomas Str\"{o}der and Terrance Swift (Editors):         Proceedings of the International Joint Workshop on Implementation of Constraint and Logic Programming Systems and Logic-based Methods in Programming Environments 2014\\
2014-14 & Florian Schmidt, Matteo Ceriotti, Niklas Hauser, and Klaus Wehrle:         HotBox: Testing Temperature Effects in Sensor Networks\\
2014-15 & Dominique G\"{u}ckel:         Synthesis of State Space Generators for Model Checking Microcontroller Code\\
2014-16 & Hongfei Fu:         Verifying Probabilistic Systems: New Algorithms and Complexity Results\\
2015-01 $^\ast$ &Fachgruppe Informatik:      Annual Report 2015\\
2015-02 & Dominik Franke:         Testing Life Cycle-related Properties of Mobile Applications\\
2015-05 & Florian Frohn, J\"{u}rgen Giesl, Jera Hensel, Cornelius Aschermann, and Thomas Str\"{o}der:         Inferring Lower Bounds for Runtime Complexity\\
2015-06 & Thomas Str\"{o}der and Wolfgang Thomas (Editors):         Proceedings of the Young Researchers' Conference ``Frontiers of Formal Methods''\\
2015-07 & Hilal Diab:         Experimental Validation and Mathematical Analysis of Cooperative Vehicles in a Platoon\\
2015-08 & Mathias Pelka, J\'{o} Agila Bitsch, Horst Hellbr\"{u}ck, and Klaus Wehrle (Editors):         Proceedings of the 1st KuVS Expert Talk on Localization\\
2015-09 & Xin Chen:         Reachability Analysis of Non-Linear Hybrid Systems Using Taylor Models\\
2015-11 & Stefan W\"{u}ller, Mari\'{a}n K\"{u}hnel, and Ulrike Meyer:         Information Hiding in the Public RSA Modulus\\
2015-12 & Christoph Matheja, Christina Jansen, and Thomas Noll:         Tree-like Grammars and Separation Logic\\
2015-13 & Andreas Polzer:         Ansatz zur variantenreichen und modellbasierten Entwicklung von eingebetteten Systemen unter Ber\"{u}cksichtigung regelungs- und softwaretechnischer Anforderungen\\
2015-14 & Niloofar Safiran and Uwe Naumann:         Symbolic vs. Algorithmic Differentiation of GSL Integration Routines\\
2016-01 $^\ast$ &Fachgruppe Informatik:      Annual Report 2016\\
2016-02 & Ibtissem Ben Makhlouf:         Comparative Evaluation and Improvement of Computational Approaches to Reachability Analysis of Linear Hybrid Systems\\
2016-03 & Florian Frohn, Matthias Naaf, Jera Hensel, Marc Brockschmidt, and J\"{u}rgen Giesl:         Lower Runtime Bounds for Integer Programs\\
2016-04 & Jera Hensel, J\"{u}rgen Giesl, Florian Frohn, and Thomas Str\"{o}der:         Proving Termination of Programs with Bitvector Arithmetic by Symbolic Execution\\

\end{longtable}
\bigskip

\noindent
{\small $^\ast$ These reports are only available as a printed version.\\
  Please contact \email{biblio@informatik.rwth-aachen.de} to obtain
  copies.}

\end{document}